\documentclass[a4paper,10pt,twoside]{cpc-hepnp}

\usepackage{multicol}
\usepackage{graphicx}
\usepackage{booktabs}
\usepackage{amssymb,bm,mathrsfs,bbm,amscd}
\usepackage[tbtags]{amsmath}
\usepackage{lastpage}

\begin{document}

\fancyhead[c]{\small Submitted to Chinese Physics C} \fancyfoot[C]{\small 010201-\thepage}

\footnotetext[0]{Received 14 March 2009}

\title{The study of neutron spectra in water bath from Pb target irradiated by 250MeV/u protons\thanks{supported by National Natural Science
Foundation and ADS special fund of China (11305229, 11105186, 91226107, 91026009, Y103020ADS) }}

\author{%
      LI Yan-Yan$^{1,2}$
\quad ZHANG Xue-Ying$^{1;1)}$\email{zhxy@impcas.ac.cn}%
\quad JU Yong-Qin$^{1}$
\quad MA Fei$^{1}$\\
\quad ZHANG Hong-Bin$^{1}$
\quad CHEN Liang$^{1}$
\quad GE Hong-Lin$^{1}$
\quad LUO Peng$^{1}$\\
\quad ZHOU Bin$^{3}$
\quad ZHANG Yan-Bin$^{1}$
\quad LI Jian-Yang$^{1}$
\quad XU Jun-Kui$^{1}$\\
\quad WANG Song-Lin$^{3}$
\quad YANG Yong-Wei$^{1}$
\quad YANG Lei$^{1}$
}
\maketitle

\address{%
$^1$ Institute of Modern Physics, Chinese Academy of Sciences, Lanzhou 730000, China\\
$^2$ University of Chinese Academy of Sciences, Beijing 100049, China\\
$^3$ Institute of High Energy Physics, Chinese Academy of Sciences, Beijing 100049, China\\
}

\begin{abstract}
The spallation neutrons were produced by the irradiation of Pb with 250 MeV protons. The Pb target was surrounded by water which was used to slow down the emitted neutrons. The moderated neutrons in the water bath were measured by using the resonance detectors of Au, Mn and In with Cd cover. According to the measured activities of the foils, the neutron flux at different resonance energy were deduced and the epithermal neutron spectra were proposed. Corresponding results calculated with the Monte Carlo code MCNPX were compared with the experimental data to check the validity of the code.
\end{abstract}

\begin{keyword}
spallation reaction, bath activation method, neutron spectrum, MCNPX simulation
\end{keyword}

\begin{pacs}
25.40.Sc, 28.20.-v, 24.10.Lx
\end{pacs}

\footnotetext[0]{\hspace*{-3mm}\raisebox{0.3ex}{$\scriptstyle\copyright$}2013
Chinese Physical Society and the Institute of High Energy Physics
of the Chinese Academy of Sciences and the Institute
of Modern Physics of the Chinese Academy of Sciences and IOP Publishing Ltd}%

\begin{multicols}{2}

\section{Introduction}

Spallation reactions can be used to produce intense neutron fluxes with a high energy proton beam on a thick target. Recently, the possible applications are rapidly growing in many fields, such as spallation neutron source(SNS) and accelerator driven system(ADS)\cite{Krivopustov,Bowman}. For designing the spallation target and shielding of accelerator facilities, it is necessary to estimate the production and distribution of the spallation reaction products especially for the neutrons. A common approach to predict the neutron production is based on the simulation with Monte-Carlo radiation transport codes, such as MCNPX \cite{MCNPX}, PHITS \cite{Niita}, GEANT4 \cite{Agostinelli} and FLUKA \cite{Battistoni}. To calculate the spallation neutrons directly produced from the target, the MCNPX  code has been verified by carrying out systematic experiment study and could give good predictions \cite{Meer}. With regard to the spatial distribution of moderated neutrons in shielding material, there are very few measurements to check the validity of the simulation code. In this work, we chose the water as the moderating material and studied the moderated energy spectra in water bath for the neutrons produced via the spallation reaction. The experimental data would be compared with the calculated results performed by MCNPX2.7.0 for testing the applicability of the code to the simulation of moderated neutron.

\section{Experimental setup}

The experiment was performed at the HIRFL-CSR in Lanzhou, China \cite{Xia}. In this experiment, we aimed at studying the moderated neutron spectra of the spallation reaction in water bath. For this purpose, a proton beam with energy of 250 MeV was used to bombard a cylindrical lead target. The beam was collimated on the centre of front surface of the target and the beam spot was less than 2 cm in diameter. The irradiation of the beam lasted about 24 hours to get sufficient fluence. In front of the target, an air ionization chamber was placed to monitor the beam current. The ionization chamber which wasn't calibrated could only show the relative amount of the beam current as function of time. The total number of protons in the process of irradiation was given by activation analysis via the yield of the reaction $^{27}$Al(p, 3p1n)$^{24}$Na. The Al-foil with high-purity and thin 800 $\mu$m was placed immediately in front of the ionization chamber. The cross section had been measured to be 10.6 mb in reference \cite{Yashima}. The total number of protons was determined to be $(1.69{\pm}0.05)\times10^{12}$. With the aid of activity analysis of Al foil, the ionization chamber was calibrated and the absolute beam current $I_{beam}$$(t)$ was obtained, see Fig. ~\ref{fig1}. The Pb target with 10 cm long by 10 cm diameter was surrounded by a tank of water. The dimensions of the container were 120 cm (L)$\times$100 cm (W)$\times$100 cm (H), where this size effectively contained most of the neutrons emitted from the target. Along the surface of the target, the activation foils of Au, Mn and In were located in the water in order to measure the moderated neutrons. The detailed description for the foils is summarized in Table.~\ref{tab1}. The foils were covered with Cd (0.8 mm thickness) to shield the thermal neutrons. All the foils were divided into three groups, and each group consisted of a piece of Au, Mn, and In foils, respectively. Three groups of foils were placed at a radial distance of 8 cm from the target axis and at the longitudinal distances of -10, 0 and 10 cm from the center of the target, see Fig.\ref{fig2}. The neutrons emitted from target were slow down in the water and measured by the resonance detectors. Via capture reaction (n,$\gamma$), the stable isotopes composing of the detector foils were transmuted into radioactive ones, which were identified by observing the characteristic $\gamma$ rays. The measurements were performed by using high-purity germanium (HPGe) detectors. The HPGe detector relative efficiency was about 65\%  and energy resolutions was 1.90 keV at 1.33 MeV. The absolute efficiency was calibrated with the standard sources $^{60}Co$, $^{133}Ba$, $^{137}Cs$, and $^{152}Eu$.

  \begin{center}
    \includegraphics[width=8cm]{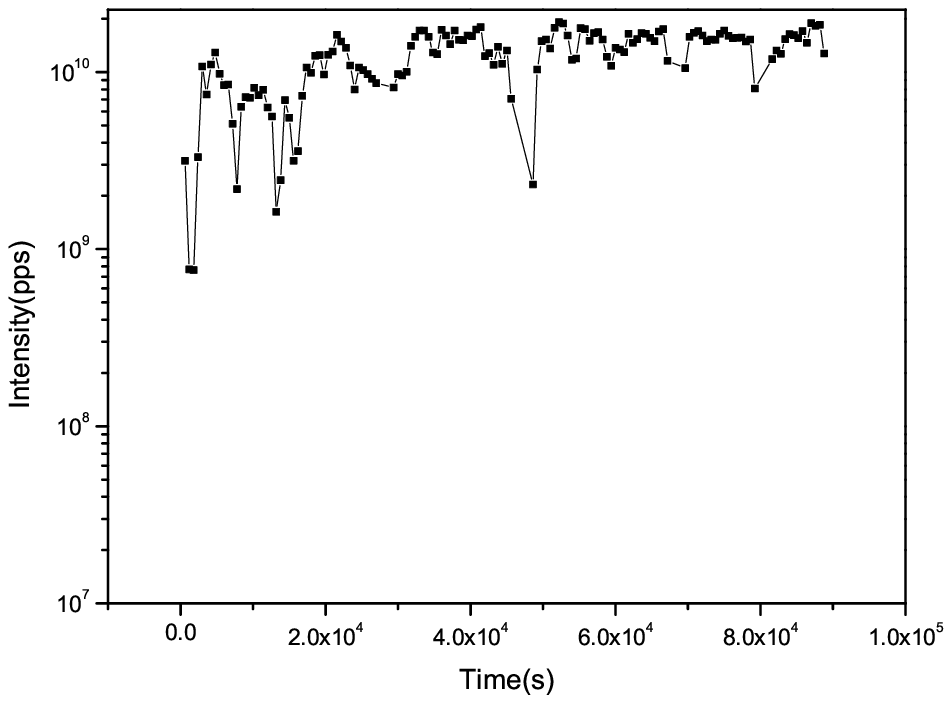}
    \figcaption{\label{fig1} Course of irradiation with 250 MeV protons.}
  \end{center}

\end{multicols}

\begin{center}
\tabcaption{ \label{tab1}  The parameters of the neutron-activation foils.}
\newcommand{\tabincell}[2]{\begin{tabular}{@{}#1@{}}#2\end{tabular}}
 {\small
  \begin{tabular}{c|c|c|c|c|c|c}
  \hline
  Activation Foils & \tabincell{c}{Area\\(mm$^2$)} & \tabincell{c}{Thickness\\(mm)} & Reaction & Half-life&
  \tabincell{c}{Energy\\(keV)} & \tabincell{c}{Branching ratio\\(\%)}\\
  \hline
  $^{197}Au$ & 20$\times$20 & 0.1 & $^{197}Au$(n,$\gamma$)$^{198}Au$ &2.69517 d& 411.8&96\\
  \hline
  $^{55}Mn$ & 20$\times$20 & 1 & $^{55}Mn$(n,$\gamma$)$^{56}Mn$ & 2.5785 h&846.8&98.9\\
  \hline
  $^{115}In$ & 20$\times$20 & 0.29 & $^{115}In$(n,$\gamma$)$^{116m}In$ & 54.29 m&\tabincell{c}{417\\ 818.7\\1097.2\\1293.5}&\tabincell{c}{28.9\\ 11.5\\56.2\\84.4}  \\
  \hline
  \end{tabular}
  }
\end{center}

\begin{center}
    \includegraphics[width=10cm]{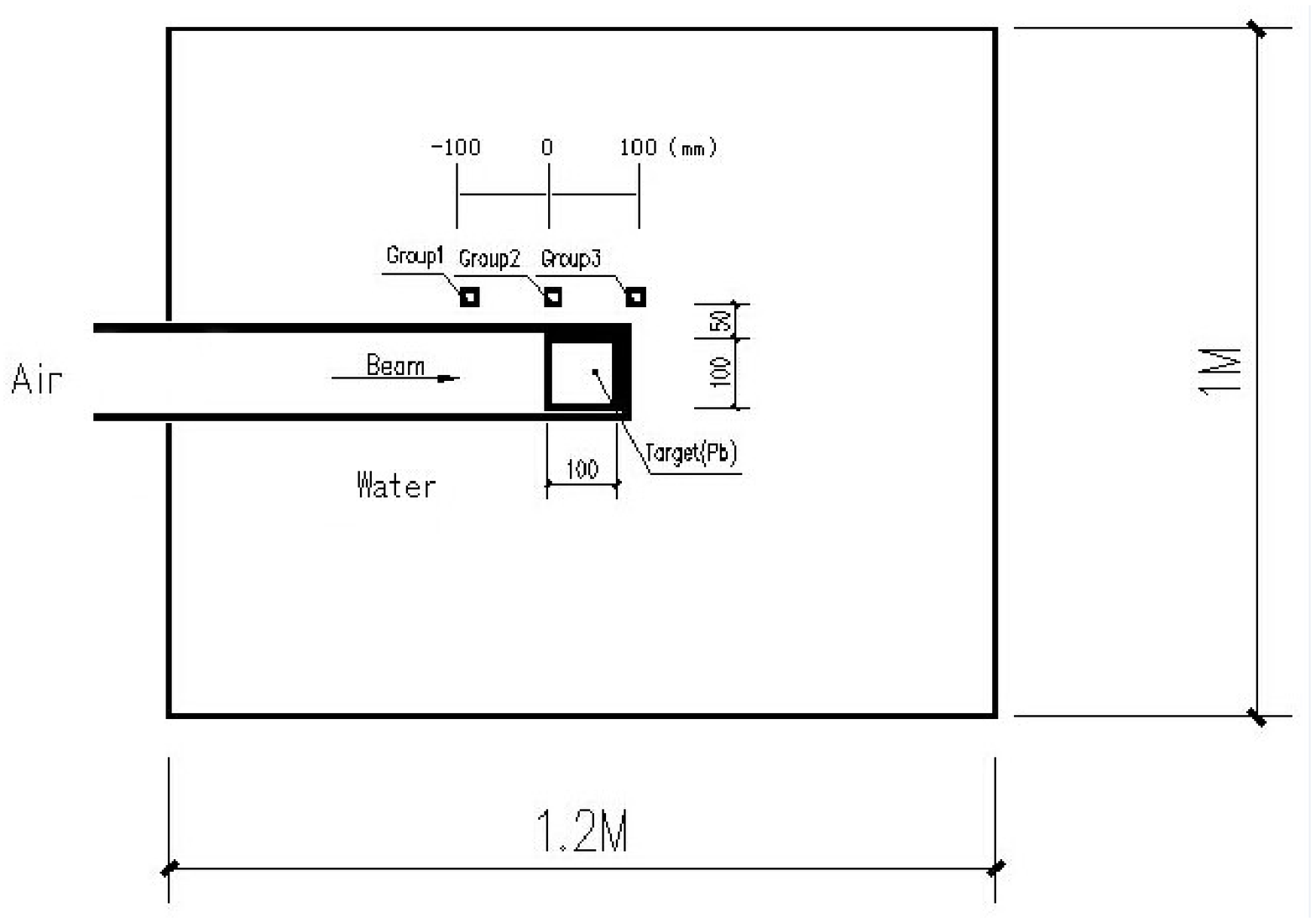}
    \figcaption{\label{fig2} The schematic view of the experimental setup.}
  \end{center}
\begin{multicols}{2}

\section{ Results and Discussion}
\subsection{ Data Analysis }

The activation analysis method is a common method to measure the neutron spectrum. The neutrons emitted from target were slowed down in the water and averaged over the Maxwell distribution of neutron energies \cite{Beckurts}. In the 1/$v$-part of the epithermal neutron field, the energy distribution can be described by

\begin{equation}
\label{eq1}
\phi(E)dE =\phi_{epi}\frac{dE}{E}.
\end{equation}

$\phi_{epi}$ is the epithermal flux per unit lethargy and does not depend on energy. The foils were irradiated in the neutron field under the sealed cadmium (Cd) cover which were used for shielding the foils from thermal neutrons. In the epithermal neutron field, each kind of the foils has a main resonance whose cross section is so large that the capture processes are responsible for the main part of the activation, see Table.~\ref{tab2} \cite{Beckurts,Yurevich}.

\end{multicols}

\begin{center}
\tabcaption{ \label{tab2}  The characteristics of the resonance detectors.}
\newcommand{\tabincell}[2]{\begin{tabular}{@{}#1@{}}#2\end{tabular}}
  {\small
  \begin{tabular}{c|c|c|c}
  \hline
  Activation Foils & \tabincell{c}{Resonance Energy\\(eV)} &  \tabincell{c}{$I_{act}=\int_{0.55eV}^{\infty}\sigma_{a}(E)/E{\rm dE}$\\(barn)} & $I_{r}/I_{act}$\\
  \hline
  $^{115}In$ & 1.457 & 2700 & $\sim$ 0.96\\
  \hline
  $^{197}Au$ & 4.905 & 1150 & $\sim$ 0.95\\
  \hline
  $^{55}Mn$ & 337 & 15.7 & $\sim$ 0.88\\
  \hline
  \end{tabular}
  }
\end{center}

\begin{multicols}{2}

 As is seen, for Au foil the main resonance is about 4.905 eV, and from the Table.~\ref{tab2}, it can be seen that the contribution of the resonance to the activation is nearly 95\% which can lead to an approximation as follows
\begin{equation}
\label{eq2}
 A_{Cd} = N \int_{E_{T}}^{\infty}\sigma_{a}(E)\phi(E){\rm dE}
 \approx N \int_{1eV}^{10eV}\sigma_{a}(E)\frac{\phi_{epi}}{E}{\rm dE}.
\end{equation}

Where $A_{Cd}$ is the activation rate (per proton). N is the atom number per gram of the foil and $\sigma_{a}(E)$ is the cross section which could be obtained from the ENDF\cite{ENDF}.
Then we can get the number of activated nuclei $N_{0}$ at the end of irradiation

\begin{equation}
\label{eq3}
N_{0}= A_{Cd}\times B_{integral}.
\end{equation}

The beam integral $B_{integral}$, corrected for the fluctuation of the proton beam current $I_{beam}(t)$, is expressed by

\begin{equation}
\label{eq4}
B_{integral}=\sum_{t=0}^{T} I_{beam}(t)\times e^{-\lambda(T-t)}.
\end{equation}

Where T is the total irradiation time, and (T-t) is the decay time of the nuclide. At the end of irradiation, we could obtain a number C of counts in a net full-energy peak of the characteristic $\gamma$-ray in a measured $\gamma$-spectrum. The activity $A_{0}$ (Bq/g) of the foil at the end of irradiation is related to the C according to the relation

\begin{equation}
\label{eq5}
A_{0}= \frac{C \lambda e^{\lambda t_{d}}}{\varepsilon_{\gamma} I_{\gamma}D K M (1-e^{-\lambda t_{c}})}.
\end{equation}
 Where $\lambda= ln2/T_{1/2}$ is the decay constant. $\varepsilon_{\gamma}$ is the efficiency of the detector and $I_{\gamma}$ is the intensity of the ¦Ã transition per decay. D and K are the correction factors for the dead time of detector and the self-absorption of ¦Ã transition in the foil, respectively. M is the mass of the foil. $t_{d}$ is time from the end of irradiation to the beginning of the measurement and $t_{c}$ is the counting time.

$A_{0}$ is related to the $N_{0}$ by the relation of  $N_{0} = A_{0}/\lambda$, so the value $A_{Cd}$  could be obtained by putting $N_{0}$ and $B_{integral}$ into the equation (3). Then we could get the neutron flux at resonance energy point by using the equation (2). For every position in the water bath, those neutron flux values of different resonance detectors can constitute a substantial spectrum. So the types of the foils are more, the spectrum is wider and more accurate.

\subsection{The Experimental Results}

Fig.~\ref{fig3} presents the Au-, Mn-, and In-foil activity data for the irradiation at the different position in the water moderator.
The activation of the foils at the -10 cm and 10 cm are both lower than the yields at 0 cm and the activation at the 10 cm is the least.  This trend allows us to conclude that the intensity of the thermal neutron flux at the position of 0 cm is higher than that at other two positions. The reason could be that when the proton beam laterally cross the lead target the neutrons generated reach maximum at the center position and the protons could reach the end of the target(10 cm) were little.
 \begin{center}
    \includegraphics[width=8cm]{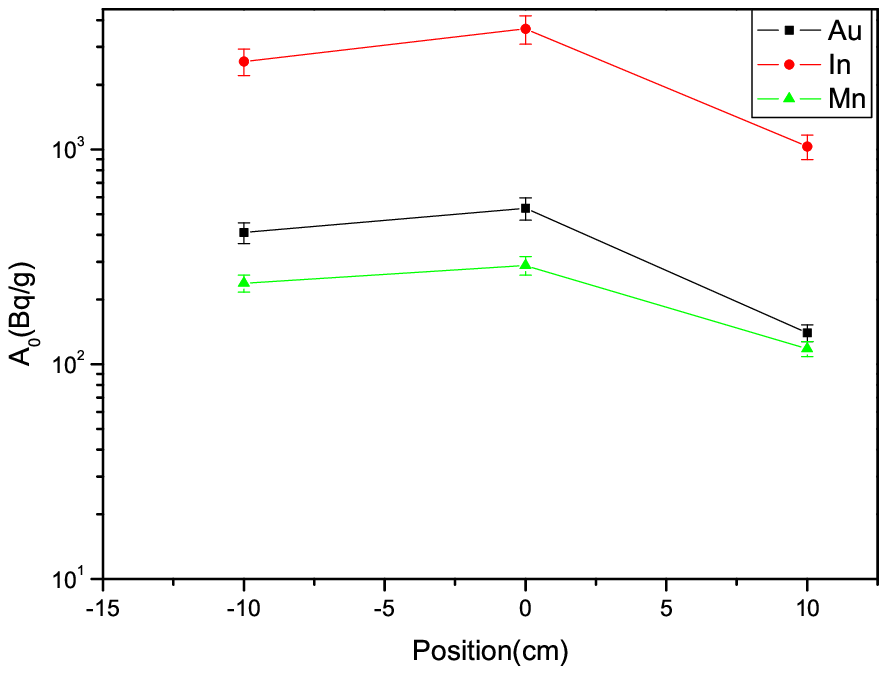}
    \figcaption{\label{fig3} (color online) Longitudinal distributions of activation yields of the foils.}
  \end{center}

According to the Eq. (3)-(5),  we can obtain the value of $A_{Cd}$ which is the activation induced by single proton, see Fig.~\ref{fig4}. The trend of $A_{Cd}$ is roughly in consistent with the value of $A_{0}$ of which the difference is mainly led by the different nuclear qualities of the foils.
  \begin{center}
    \includegraphics[width=8cm]{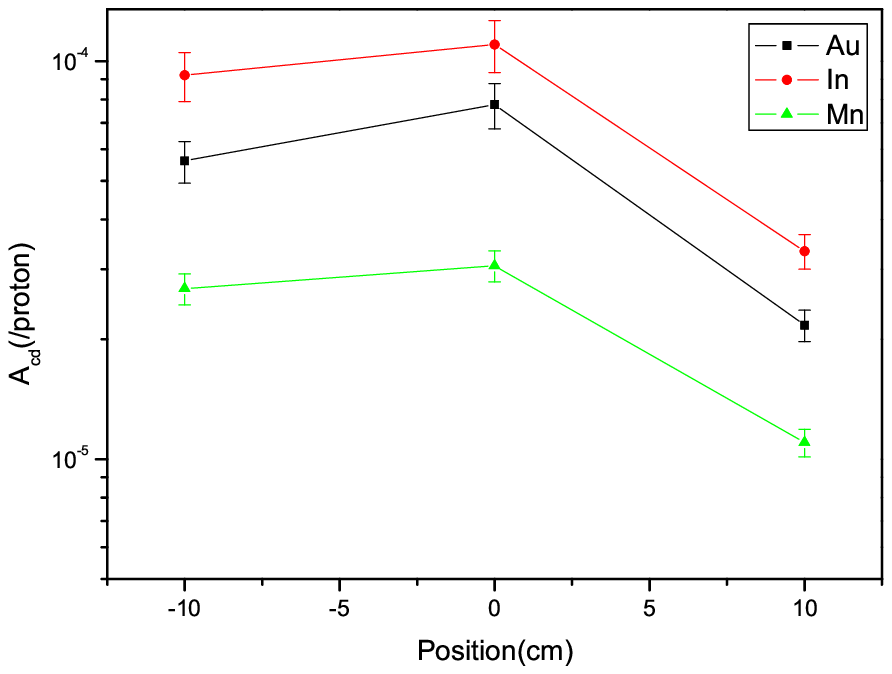}
    \figcaption{\label{fig4} (color online) Activation rate of the foils along the longitudinal axis.}
  \end{center}

Then combining the equation (2), we can get the $\phi_{epi}$ which is the epithermal flux per unit lethargy. The neutron flux at the main resonance energy points of Au, Mn and In could be deduced from the equation (1). According to the neutron flux in different positions of water bath, we can obtain the approximate spectra in the main resonance energy range of Au, Mn and In, see Fig.~\ref{fig5}. It could be seen that the neutron flux at the center of the target(0 cm) has the biggest value and at the end of the target(10 cm) the flux is the least.

 \begin{center}
    \includegraphics[width=8cm]{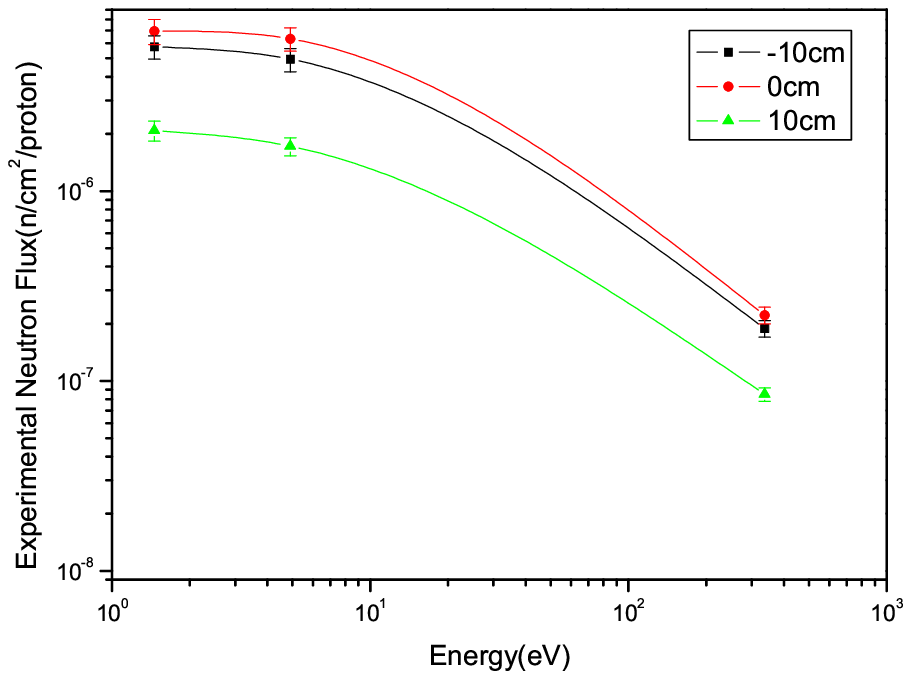}
    \figcaption{\label{fig5} (color online) The measured neutron spectra in different positions of water bath.}
  \end{center}

\subsection{Simulation and Comparison}

The process of the irradiation experiment was accompanied with the simulation by using the MCNPX2.7.0. In the intra-nuclear cascade (INC) stage, we chose the Bertini model \cite{Bertini} for nucleons and ISABEL model \cite{Yariv} for other particle types. The pre-equilibrium model was used after INC stage. The RAL model \cite{Atchison} was used in the process of evaporation (or fission).

The comparison of $A_{0}$ and $A_{Cd}$ between the simulation and experiment is shown in Fig.~\ref{fig6} and Fig.~\ref{fig7}. From Fig.~\ref{fig6} and Fig.~\ref{fig7}, it can be seen that measured and calculated values of Au foils comply better, where the maximum difference is less than 15$\%$. For Mn foils, Mn-2 (in group2) in the center place comply better. But both sides including Mn-1 (in group1) and Mn-3 (in group3) are with big errors. The reason for this could be the finite thickness of Mn foil (1 mm) whose resonance is self-shielded and indeed more strongly the higher the resonance given by Table.~\ref{tab2}. For In foils, the measured results are some 10$\%$ to 30$\%$ lower than the calculated value. This implied that the In capture cross-sections are less precisely than the corresponding ones for Au. Thus, only the $^{197}Au$ activity data are good agreement with the calculated results.

 \begin{center}
    \includegraphics[width=8cm]{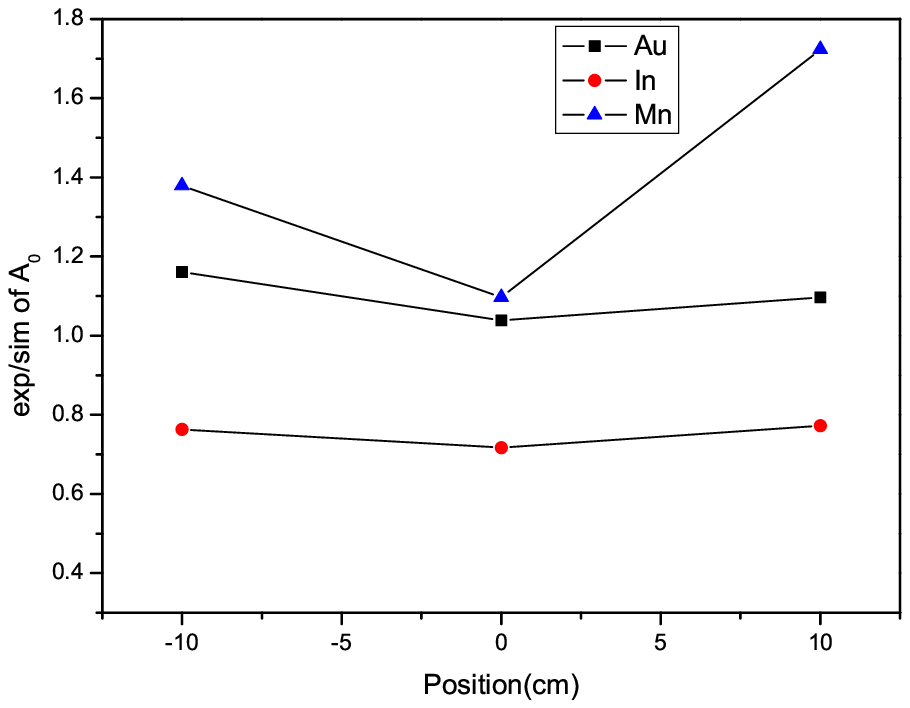}
    \figcaption{\label{fig6} (color online) Comparison of the experimental $A_{0}$ versus the $A_{0}$ from the MCNPX simulation.}
  \end{center}

 \begin{center}
    \includegraphics[width=8cm]{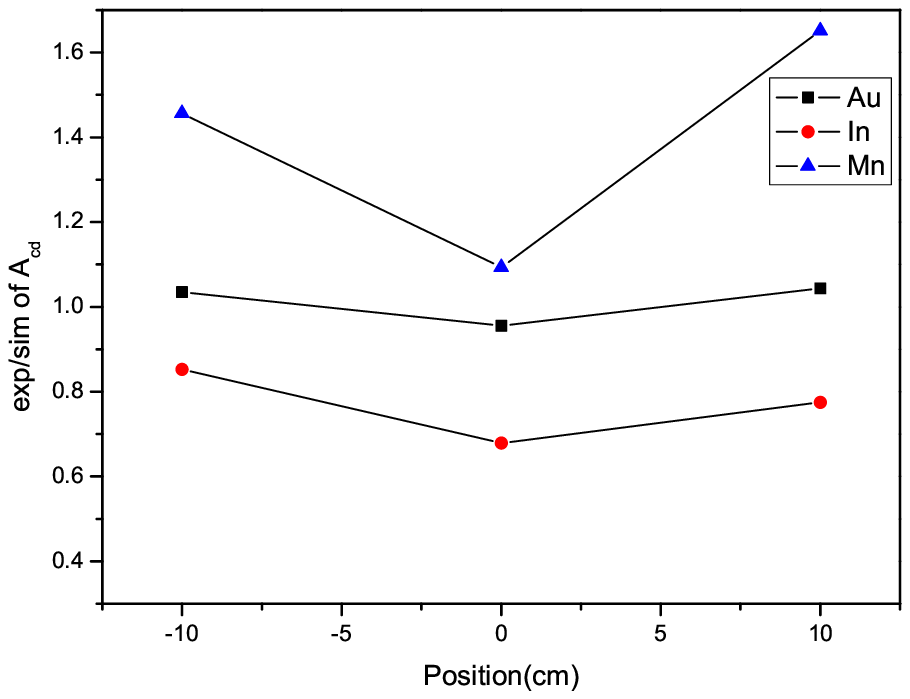}
    \figcaption{\label{fig7} (color online) Comparison of the experimental $A_{Cd}$ versus the $A_{Cd}$ from the MCNPX simulation.}
  \end{center}

In order to investigate the cause of the discrepancy between the calculations and the experiments, we simulated the energy spectra of all the three positions where the foils were located in the water. The calculated results with MCNPX code system are compared with the present experimental results in Fig.~\ref{fig8}. The purple line is made of the experimental points and the other three lines are the simulated neutron spectrum in the whole energy range. It is observed that the calculated results are in good agreement with experimental spectrum above 50 eV. For neutrons below 50 eV, the MCNPX calculations give about two times or more higher flux than the experiment. At the same time, it can be seen that in simulated spectrum there are some sudden drops at the resonance energy points of the foils. This situation could be attributed to the foils themselves, since many neutrons with the resonance energy were absorbed by the foils and not recorded into the energy spectra. This indicated that the foils absorbed so many resonance neutrons that the neutron fields in the water bath were badly effected.

 \begin{center}
    \includegraphics[width=6cm]{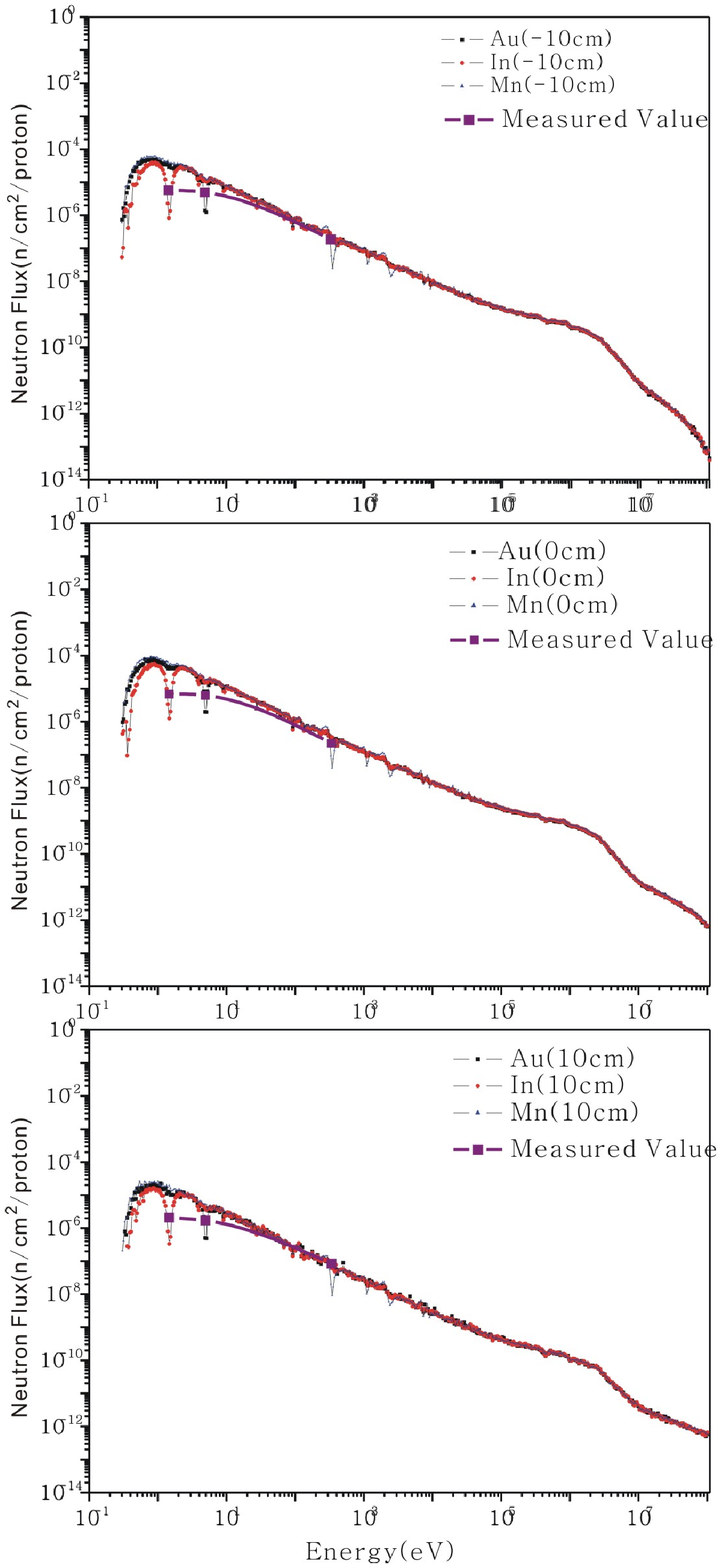}
    \figcaption{\label{fig8} (color online) Comparison between the measured spectra and the calculated spectrum.}
  \end{center}

Taking into account the impact of foils on the simulation, we also simulated a model which was the same with the model in Fig.~\ref{fig8} except there were no foils around the target. We just wanted to know whether the foils had  any appreciable or noticeable impact on the neutron spectra. As shown in Fig.~\ref{fig9}, we can see that when the foils were not represented in the simulation the neutron energy spectra had the smooth and continuous curves. But when foils were included the flux have a sudden drop at the resonance energy points, see Fig.~\ref{fig8}. Excluding the neutron flux at the resonance energy points, there is no obvious difference between Fig.~\ref{fig8} and Fig.~\ref{fig9}.

 \begin{center}
    \includegraphics[width=6cm]{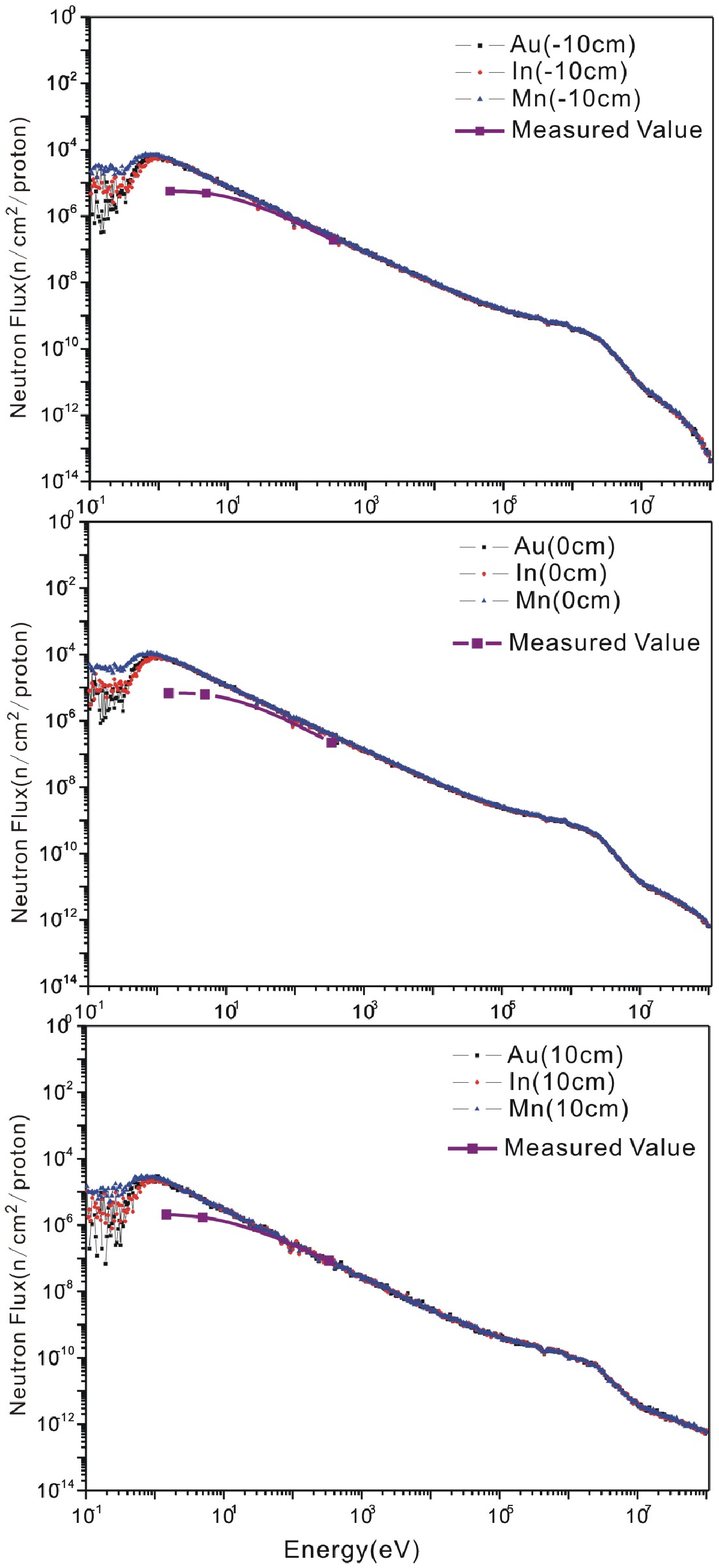}
    \figcaption{\label{fig9} (color online) Comparison between the measured spectra and the calculated spectrum without foils.}
  \end{center}

\section{Conclusions}

In summary, we have studied the neutron spectra in specific energy range via the reaction by using protons bombarding a thick lead target in water bath. The intensity of the beam and the activation of the resonance detectors were measured by the neutron activation analysis method. By analyzing the activation of different foils in different positions near the lead target, the neutron fluxes were obtained and the corresponding neutron spectra were given. We also compared the experiment data with the simulations. It was found that the calculations were in agreement with the experimental data in high energy range, and it should be emphasized that the simulations must value the contribution of detectors themselves. This is just the first step in the neutron spectrum measurement. And next, we will try to use more resonance detectors of which the resonance energy is different with wider distribution.

\section{Acknowledgments}

We thank the support of the accelerator operation staff at HIRFL-CSR. 
\end{multicols}

\end{document}